\documentclass[aps,
showpacs,prl
,preprint
]{revtex4}
\usepackage{epsf}
\begin{document}
\title{Two-State Folding, Folding Through Intermediates, and Metastability 
in a Minimalistic Hydrophobic-Polar Model for Proteins}
\author{Stefan Schnabel}
\email[E-mail: ]{Stefan.Schnabel@itp.uni-leipzig.de}
\author{Michael Bachmann}
\email[E-mail: ]{Michael.Bachmann@itp.uni-leipzig.de}
\author{Wolfhard Janke}
\email[E-mail: ]{Wolfhard.Janke@itp.uni-leipzig.de}
\homepage[\\ Homepage: ]{http://www.physik.uni-leipzig.de/CQT}
\affiliation{Institut f\"ur Theoretische Physik and Centre for Theoretical Sciences (NTZ), 
Universit\"at Leipzig, Augustusplatz 10/11, D-04109 Leipzig, Germany}
\begin{abstract}
Within the frame of an effective, coarse-grained hydrophobic-polar protein model, we employ
multicanonical Monte Carlo simulations to investigate free-energy landscapes and 
folding channels of exemplified heteropolymer sequences, which are permutations of each other. 
Despite the simplicity of the model, the knowledge of the free-energy 
landscape in dependence of a suitable system order parameter enables us to reveal complex
folding characteristics known from real bioproteins and synthetic peptides, such as two-state folding, 
folding through weakly stable intermediates, and glassy metastability. 
\end{abstract}
\maketitle
Folding of linear chains of amino acids, i.e., bioproteins and synthetic peptides, is, 
for single-domain macromolecules, accompanied by the formation of secondary structures 
(helices, sheets, turns) and the tertiary hydrophobic-core collapse. While secondary
structures are typically localized to segments of the peptide, the effective hydrophobic 
interaction between nonbonded, nonpolar amino acid side chains results in a global,
cooperative arrangement favoring folds with compact hydrophobic core and surrounding polar
shell screening the core from the polar solvent. 
Systematic analyses for unravelling general folding principles are extremely difficult in microscopic all-atom approaches, since the folding process is 
strongly dependent on the ``disordered'' sequence of amino acids~-- twenty different types 
can typically occur in bioproteins~-- and the native-fold formation is inevitably connected with, 
at least, significant parts of the sequence. Moreover,
for most proteins, the folding process is relatively slow (microseconds to seconds), which is  
due to a complex, rugged shape of the free-energy landscape~\cite{onuchic0,clementi1,onuchic1} with 
``hidden'' barriers, depending on sequence properties. 
Although there is no obvious system parameter that allows for a general 
description of the accompanying conformational transitions in folding processes 
(as, for example, the reaction coordinate in chemical reactions),
it is known that there are only a few classes of characteristic folding behaviors, mainly single-exponential
folding, two-state folding, folding through intermediates, and glass-like folding into metastable
conformations~\cite{du1,pande2,okamoto1,wolynes1,pande1,pitard1}. 

An important step forward towards a better theoretical understanding of the basic
mechanisms underlying these different classes could be the design and analysis of suitably
designed coarse-grained models focusing on mesoscopic scales.
The idea to use a strongly simplified model is two-fold:
Firstly, it is believed that tertiary folding is mainly based on effective hydrophobic interactions
such that atomic details play a minor role. Secondly, systematic comparative folding studies for mutated 
or permuted sequences are computationally extremely demanding at the atomic level and are to date virtually
impossible for realistic proteins.
In this Letter, we show that by employing a coarse-grained hydrophobic-polar heteropolymer model~\cite{still1} 
and monitoring a simple angular ``order'' parameter it is indeed possible to identify 
different complex folding characteristics. 
This is comparable to studies of phase transitions based on effective order parameters in other 
disordered systems such as, e.g., spin glasses, where simplified models are successfully employed~\cite{sglass}.
The individual folding trajectories as discussed in this work will be characterized by
a similarity parameter which is related to the replica overlap parameter used in spin-glass analyses.
This is useful as the amino acid sequence induces intrinsic disorder and frustration into the system and
therefore a peptide behaves similar to a spin system with a quenched disorder configuration of couplings.

The simplified model~\cite{still1} used incorporates only two types of 
amino acids, hydrophobic and polar residues~\cite{dill1}, and focuses on qualitative aspects 
of tertiary heteropolymer folding, such as hydrophobic-core 
formation~\cite{sorenson1,bj0,hsu1,liang1,baj2}. 
This physical, effective-potential approach has to be distinguished from knowledge-based
models~-- typically of G\={o} type~-- where the contact map
of the final fold already enters as input into the model. The latter models have proven to be useful 
in understanding two-state folding of selected proteins~\cite{clementi2,li1,koga1,kaya1,schonbrun1,head1}.  
On the other hand, the kinetics of effective models is not biased towards 
a given structure, and a variety of folding behaviors can be studied. This has particular 
implications for non-two-state folding and
metastability, the latter primary concerning designed synthetic peptides or mutated biopolymers.

Our results are obtained by employing the standard hydrophobic-polar off-lattice AB model~\cite{still1} in three dimensions
for the three sequences listed in Table~\ref{tab:seqs}. The sequences were chosen from the set of deliberately designed 
sequences in Ref.~\cite{irb1} and have the same content of hydrophobic $A$ (14 each) and polar $B$ (6 each) residues.
In the following, we denote by ${\bf r}_i$ the spatial position of the $i$th monomer in the chain 
${\bf X}=\{{\bf r}_1,\ldots,{\bf r}_N\}$ of $N$ residues. Covalent bonds have unit length. The bending 
angle between monomers $k$, $k+1$, and $k+2$ is $\vartheta_k$ 
($0\le \vartheta_k\le \pi$) and $\sigma_i=A,B$ symbolizes the type of the monomer. The energy of a 
conformation is given by $E=E_{\rm bend}+E_{\rm LJ}$, where 
\begin{equation}
E_{\rm bend} =\frac{1}{4}\sum_k(1-\cos \vartheta_k)
\end{equation}
is the bending energy and 
\begin{equation}
E_{\rm LJ}=4\sum_{j>i+1}\left(r_{ij}^{-12}-C(\sigma_i,\sigma_j)r_{ij}^{-6} \right)
\end{equation}
is the contribution of the residue-type dependent Lennard-Jones potential, which depends on the distance 
$r_{ij}$ of all pairs of nonbonded monomers $i$ and $j$, being
long-range attractive for $AA$ and $BB$ pairs [$C(A,A)=1$, $C(B,B)=0.5$] and 
repulsive for $AB$ pairs of monomers [$C(A,B)=C(B,A)=-0.5$].
\begin{table}
\caption{\label{tab:seqs} The three AB 20-mers studied in this Letter
and the values of the associated (putative) global energy minima. Note that the given values for sequence S3
belong to two different, almost degenerate folds.}
\begin{tabular}{cp{3mm}cc}\hline\hline
label & & sequence & global energy minimum\\ \hline
S1 & & $BA_6BA_4BA_2BA_2B_2$ & $-33.8236$\\ 
S2 & & $A_4BA_2BABA_2B_2A_3BA_2$ & $-34.4892$\\ 
S3 & & $A_4B_2A_4BA_2BA_3B_2A$ & $-33.5838$, $-33.5116$ \\ \hline \hline
\end{tabular}
\vspace*{-3mm}
\end{table}
Simulations of this model were performed 
using standard multicanonical Monte Carlo techniques~\cite{muca1} with spherical updates~\cite{baj2}. 
For each sequence, 10 independent simulations were performed and a total statistics
of $2\times 10^9$ conformations entered into the data analysis.

Since the number of degrees of freedom (virtual bond and torsion angles) in the coarse-grained model 
is comparable with the number of dihedral angles in all-atom protein models, AB heteropolymer folding 
is of similar complexity. The main advantage is the drastically reduced computational effort
for calculating the interactions, which allows more comprehensive and systematic analyses of free-energy landscapes and folding channels
in comparative studies for different sequences. In the following, we perform such an analysis
of characteristic folding behaviors based on a suitably defined generalized angular overlap parameter, as 
introduced in Ref.~\cite{baj2} in analogy to all-atom 
studies~\cite{okamoto1}. It is a computationally low-cost measure for the similarity of two conformations,
where the differences of the angular degrees of freedom are calculated. In order to consider this 
parameter as kind of order parameter, it is useful to compare conformations ${\bf X}$ 
of the actual ensemble with
a suitable reference structure ${\bf X}^{(0)}$, which is preferably chosen to be the global-energy minimum 
conformation. The overlap parameter is defined as~\cite{baj2}
\begin{equation}
Q({\bf X})=1-d({\bf X}).
\end{equation}
Denoting by \mbox{$N_b=N-2$} and \mbox{$N_t=N-3$} the numbers of 
bending angles $\vartheta_i$ and torsional angles $\varphi_i$, respectively,
the angular deviation between the conformations is calculated according to
$d({\bf X})=\left[\sum_{i=1}^{N_b}d_b(\vartheta_i)+
{\max}\left(\sum_{i=1}^{N_t}d_t^{(+)}(\varphi_i),\sum_{i=1}^{N_t}d_t^{(-)}(\varphi_i)\right)\right]/\pi(N_b+N_t),$
where
$d_b(\vartheta_i)=|\vartheta_i-\vartheta^{(0)}_i|$ and
$d_t^{(\pm)}(\varphi_i)={\rm min} (|\varphi_i\pm\varphi^{(0)}_i|,2\pi-|\varphi_i\pm\varphi^{(0)}_i|)$.
Note that this expression takes into account the reflection symmetry $\varphi_i\to-\varphi_i$
of the AB model. Reflection-symmetric conformations are not distinguished and therefore 
only the larger overlap is considered.
The overlap is unity, if all angles coincide, else $0\le Q<1$. The average 
overlap of a random conformation with the reference state is for the three sequences close to 
$\langle Q\rangle= 0.66\pm 0.02$. Significant similarity is typically found if $Q>0.8$. 

\begin{figure*}
\centerline{\epsfxsize=16.6cm \epsfbox{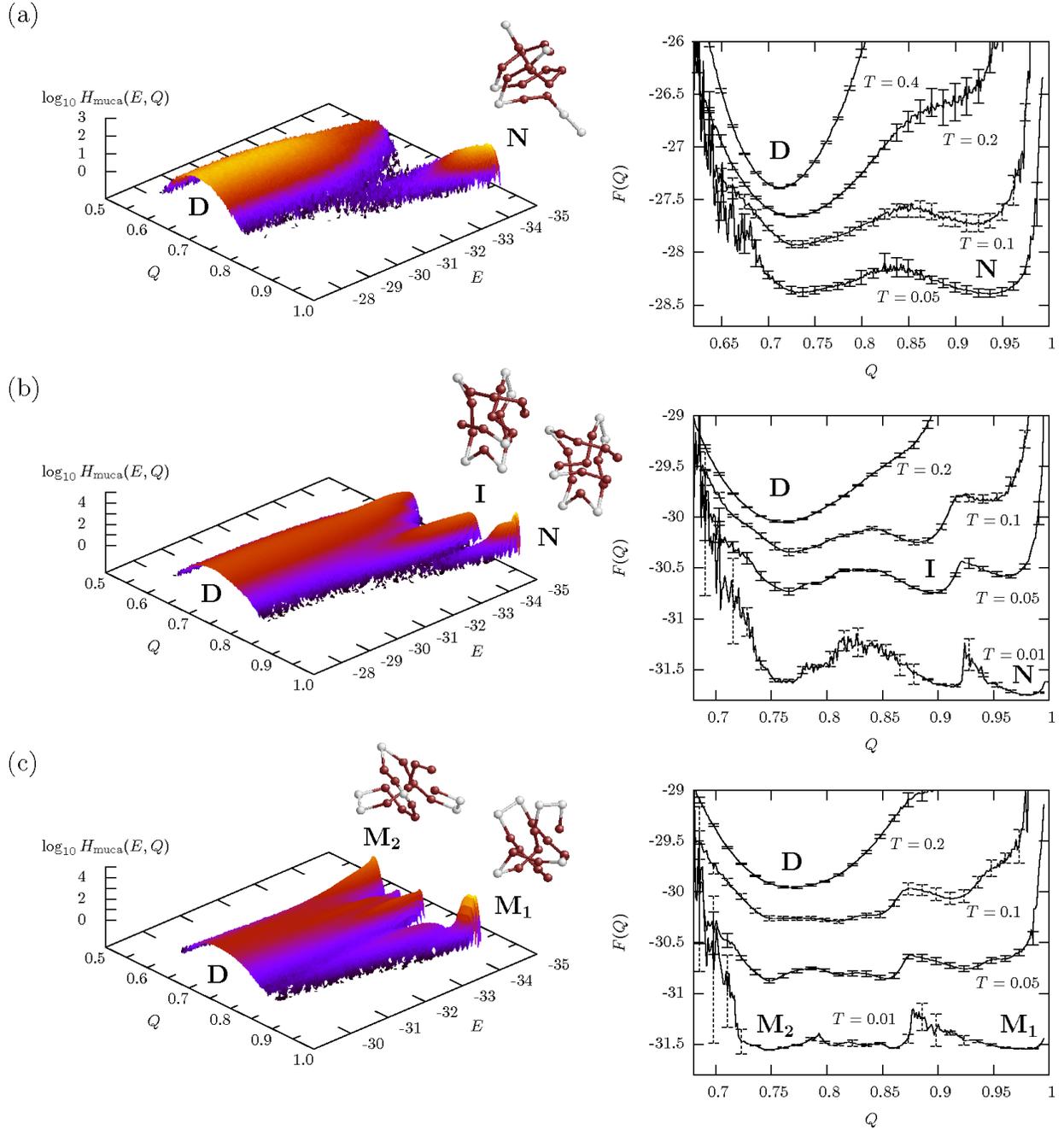}}
\caption{\label{fig:fandh} (Color online) Multicanonical histograms $H_{\rm muca}(E,Q)$ of energy $E$ and angular overlap parameter $Q$ 
and free-energy landscapes $F(Q)$ at different temperatures for the three sequences (a) S1, (b) S2, and (c) S3. 
The reference folds reside at $Q=1$ and $E=E_{\rm min}$. Pseudophases are symbolized by D (denatured states), N (native
folds), I (intermediates), and M (metastable states). Representative conformations in intermediate and folded phases
are also shown.
\vspace*{-5mm}}
\end{figure*}
For the qualitative discussion of the folding characteristics, we 
consider the multicanonical histograms of energy $E$ and angular overlap $Q$,
$H_{\rm muca}(E,Q) = \sum_t\,\delta_{E,E({\bf X}_t)}\delta_{Q,Q({\bf X}_t)}$,  
where the sum runs over all Monte Carlo sweeps $t$ in the multicanonical simulation, which yields
a constant energy distribution $h_{\rm muca}(E)= \int_0^1dQ\,H_{\rm muca}(E,Q) \approx {\rm const.}$
In consequence, $H_{\rm muca}(E,Q)$ is useful for identifying the folding channels, 
independently of temperature. Restricting the canonical partition function at temperature $T$ to the ``microoverlap'' ensemble
with overlap $Q$, $Z(Q)= \int {\cal D}{\bf X}\, \delta(Q-Q({\bf X}))\,\exp\{-E({\bf X})/k_BT\}$,
where the integral is over all possible conformations ${\bf X}$, we define the overlap free energy 
as $F(Q)=-k_BT\ln Z(Q)$. 

Figures~\ref{fig:fandh}(a)--(c) show the thus obtained multicanonical histograms $H_{\rm muca}(E,Q)$ (left) 
and the overlap free-energy landscapes $F(Q)$ (right) at different temperatures for the three sequences listed in 
Table~\ref{tab:seqs}. The different branches of $H_{\rm muca}(E,Q)$ indicate the channels the heteropolymer 
can follow in the folding process towards the reference structure. 
The heteropolymers, whose sequences differ only by permutations, exhibit noticeable differences in the folding behavior 
towards the native conformations. The first interesting observation is that the minimalistic model used is
capable of revealing the different folding behaviors of the wild-type and permuted sequences. The second remarkable
result is that the angular overlap parameter $Q$ is a surprisingly manifest measure for the peptide macrostate.

From Fig.~\ref{fig:fandh}(a) we conclude that folding of sequence S1 exhibits a typical
two-state characteristics. Above the transition, conformations 
possess a random-coil-like overlap $Q\approx 0.7$, i.e, there is no significant 
similarity with the reference structure. Close to $T\approx 0.1$ the global minimum of the corresponding 
overlap free energy $F(Q)$ changes discontinuously towards larger $Q$ values, and 
at the transition state the denatured (D) and the folded native (N) macrostate are equally probable. The existence of 
this pronounced transition state is a characteristic indication for first-order-like two-state folding. 
Decreasing the temperature further, the native-fold-like conformations ($Q>0.95$) dominate and fold smoothly 
towards the $Q=1$ reference structure, i.e., the lowest-energy conformation (N) found for sequence S1, which is also
depicted in Fig.~\ref{fig:fandh}(a).

The folding behavior of sequence S2 is significantly different, as Fig.~\ref{fig:fandh}(b) shows, and
is a typical example for a folding event through an intermediate (I) macrostate. The main channel (D)
bifurcates and a side channel (I) branches off continuously. For smaller energies (or lower temperatures), 
this branching is followed by the formation 
of a third channel, which ends in the native fold (N). The characteristics of folding-through-intermediates
is also reflected by the free-energy landscapes. Starting at high temperatures in
the pseudophase D of denatured conformations ($Q\approx 0.76$),
the intermediary phase I with $Q \approx 0.9$ is reached close to the temperature $T\approx 0.05$.
Decreasing the temperature further below the native-folding
threshold close to $T=0.01$, the hydrophobic-core formation is finished and stable native-fold-like conformations N with
$Q>0.97$ dominate.

The most extreme behavior of the three exemplified sequences is found for sequence S3, where
the main channel (D) does not decay in favor of a native-fold channel. In fact, in Fig.~\ref{fig:fandh}(c) 
we observe both, {\em two} separate native-fold channels (M$_1$ and M$_2$) and the main channel. 
Above the folding transition ($T=0.2$), the typical 
sequence-independent denatured (D) conformations ($Q\approx 0.77$) dominate. 
Annealing below the glass-transition threshold, 
several channels form and coexist. The two most prominent channels (to which the lowest-energy
conformations M$_1$ and M$_2$ belong that we found in the simulations) 
eventually lead for $T\approx 0.01$ to ensembles of states M$_1$
with $Q>0.97$, which are similar to the reference structure shown, and 
conformations M$_2$ with $Q\approx 0.75$. The lowest-energy conformation found in M$_2$ is also shown in 
Fig.~\ref{fig:fandh}(c). It is structurally different but energetically almost degenerate compared with the 
reference structure. 
It should also be noted that the lowest-energy main-channel conformations
have only slightly larger energies than the two native folds. Thus, the folding of this heteropolymer
is accompanied by a very complex, amorphous folding characteristics. In fact, the multiple-peaked distribution $H_{\rm muca}(E,Q)$
near minimum energies is a strong indication for metastability and bears similarities with 
spin-glass characteristics. A native fold in the natural sense
does not exist, the $Q=1$ conformation is only a reference structure but the folding towards this
structure is not distinguished as it is in the folding characteristics of sequences S1 and S2.

We have confirmed our results of the angular overlap analysis for the folding behaviors 
by a corresponding study of the root mean square deviation (rmsd) which is frequently used to characterize 
folding trajectories in free-energy landscapes. The main advantage of using our angular overlap parameter is its 
efficient calculation
which leads to a speed-up of computing time by a factor of about 10 compared with the efforts
required for analysing the folding channels based on the rmsd~\cite{ssbj1}.

To summarize, we have demonstrated in this study that within a minimalistic heteropolymer frame 
it is possible to find clear indications for three different 
folding characteristics known from real proteins by analysing macrostates based on an angular overlap
parameter. 
Our primary physical objective is a more comprehensive, qualitative understanding of universal
aspects of tertiary protein folding, where microscopic details are expected to be of less relevance
and which are, therefore, averaged out at a mesoscopic scale in a coarse-grained model. 
For selected hydrophobic-polar
heteropolymer sequences~-- not being explicitly designed for this study~-- we have shown that
characteristic folding behaviors such as two-state folding, folding through intermediates, 
and metastability can be identified which are qualitatively comparable with real folding events in nature.
Beyond the general interest in a theoretical understanding of the basic mechanisms of protein folding, the preparation
of synthetic peptide macrostates in future applications, e.g., the successful design of substrate- or pattern-selective 
polymers~\cite{belcher1,schulten1,goede1,bogner1,bj2}, is strongly connected with the complex aspects of conformational folding transitions
as investigated in this study.

This work is partially supported by the DFG (German Science Foundation) under Grant  
No.\ JA 483/24-1. Some simulations were performed on the 
supercomputer JUMP of the John von Neumann Institute for Computing (NIC), Forschungszentrum
J\"ulich under Grant No.\ hlz11. 
\end{document}